\documentstyle[12pt]{article}

\newcommand{\bmat}{\left(\begin{array}}
\newcommand{\emat}{\end{array}\right)}
\def\NPB#1#2#3{Nucl. Phys. B{#1} (19#2) #3}
\def\PLB#1#2#3{Phys. Lett. B{#1} (19#2) #3}

\def\yzero{\smash{\hbox{$y\kern-4pt\raise1pt\hbox{${}^\circ$}$}}}

\def\a{\alpha}

\def\-{\hphantom{-}}
\def\ov{\overline}
\def\s2{\frac{1}{2}}

\def\oh{\frac{1}{2}}
\def\beq{\begin{equation}}
\def\eeq{\end{equation}}
\def\beqa{\begin{eqnarray}}
\def\eeqa{\end{eqnarray}}

\def\diag{{\rm diag \,}}
\def\IF{\relax{\rm I\kern-.18em F}}
\def\II{\relax{\rm I\kern-.18em I}}
\def\IP{\relax{\rm I\kern-.18em P}}
\def\IC{\relax\hbox{\kern.25em$\inbar\kern-.3em{\rm C}$}}
\def\IR{\relax{\rm I\kern-.18em R}}

\def\Dsl{\,\raise.15ex\hbox{/}\mkern-13.5mu D} 
\def\IZ{\bf Z}
\def\IC{\bf C}
\def\id{{\rm I}}

\topmargin
-1.5cm
\textwidth
15.5cm
\textheight
23.5cm
\oddsidemargin
0.7cm
\evensidemargin
1.2cm

\begin{document}
\pagestyle{empty}
\rightline{CAB-IB/ 2900802, FTUAM-02/7 IFT-UAM-02/6,SISSA Ref. 19/2002/EP}
\rightline{\tt hep-th/0203129}
\vspace{0.5cm}
\begin{center}
\LARGE{In  quest of  ``just'' the Standard Model on 
D-branes at a singularity\\[10mm]}\medskip
\large{L. F.~ Alday $^{1,2}$ and G.~Aldazabal $^{3,4}$
\\[2mm]}
\small{ $^1$ Sissa, Triste, Italy \\[3mm]
$^2$ The Abdus Salam ICTP, Triste, Italy \\[3mm]
$^3$  Departamento de F\'{\i}sica Te\'orica C-XI
and Instituto de F\'{\i}sica Te\'orica  C-XVI,\\[-0.3em]
Universidad Aut\'onoma de Madrid,
Cantoblanco, 28049 Madrid, Spain \\[3mm]
$^4$  Instituto Balseiro and Centro At\'omico
Bariloche, \\
8400 S.C. de Bariloche, (CNEA  and CONICET), Argentina
\\[9mm]}
\small{\bf Abstract} \\[7mm]
\end{center}
\begin{center}
\begin{minipage}[h]{14.0cm}
In this note we explore  the possibility of obtaining  gauge
bosons and fermionic spectrum as close as possible to the Standard
Model content, by placing D3-branes at a $\IZ_N$ orbifold-like
singularity in the presence of D7-branes. Indeed, we find
that this is plausible provided  a sufficiently high $N$ is allowed for and
 the singular point is also fixed by an orientifold action. If 
extra charged matter is not permitted then the
singularity should  necessarily be non-supersymmetric. 
Correct hypercharge assignments require a dependence on some Abelian 
gauge D7-groups.
In achieving such a construction we follow a recent observation made in
Ref.\cite{imr} about the possibility that,  the three left handed
quarks, would  present  different $U(2)$ transformation properties.
\end{minipage}
\end{center}

\newpage
We encode under the name of {\it String phenomenology} \cite{pheno} the
different attempts to embed the Standard Model of fundamental
interactions, or plausible extensions of it,  into the framework
of string theory.  Activity in this area started in the middle eighties, 
especially in  the so called
{\it perturbative} heterotic string context, and 
many features have been understood since then. 
A lesson to recall is  that, in spite of the enormous degeneracy of $D=4$
dimensional  string vacua, leading to loss of
predictability, not everything can  be fitted into such a context.
 String theory imposes  severe constraints indeed on model
building.
 A neat example is provided, for instance,  by the fact that, in heterotic 
string theory,  the
contribution to the mass of a state in a given representation of the
gauge group is proportional to
the dimension of the representation. Thus, high dimensional massless 
representations are not allowed for in perturbative heterotic string models
\footnote{This appears to be an important limitation to GUT like
models \cite{guts}}.

 Stringy constraints, related to the structure of anomaly cancellation, 
can also be found behind the {\it failure} to build  exactly the Standard 
Model (SM) in a string theory framework.
 Despite the  many models constructed with gauge and
massless fermionic sector quite close to the SM one,  such models
generically  possess extra visible fermionic matter.  This is valid
for heterotic perturbative constructions and  also for other
string constructions involving, for instance,  Type II  D-branes.

 This may seem  rather  surprising since Standard Model, 
$SU(3)\times SU(2)\times U(1)_Y$,  generations already produce anomaly 
free combinations. 
String consistency requirements, like  
modular invariance  in perturbative heterotic string  or tadpole cancellation 
in open string models, certainly imply that  such models are free of 
anomalies. However, we should notice that stringy constraints are generically 
stronger than anomaly cancellation conditions. 
 These stronger requirements often manifest in the presence of extra gauge 
factors,  and  thus, of extra chiral fermions which 
must be generically present for canceling their  anomalies.
For instance, in D-brane models, $U(2)$ unitary groups appear, rather than 
$SU(2)$. As a consequence,  doublets $(2_1)$
or anti-doublets $ 2_{-1}$   are distinguished by their different $U(1)$ 
charges  (indicated as  1 and  -1 subscripts   respectively) 
and tadpole cancellation requires the same number of both of them.
  
Hence, if the three  left handed quark   
generations were just mere replications of each other, say $ 3 (3,2)$,  then 
9 anti-doublets should  also be present. 
Given that  three  of them can  be identified with SM leptons,  
still six extra doublets will be required by stringy considerations. 
Nevertheless, if two left 
quarks were doublets (or anti-doublets) and the other one was an $U(2)$ 
anti-doublet (doublet)  then no extra doublets would be needed.     


It is by noticing this fact that, only very recently \cite{imr},  string
models  with ``just''  the, non supersymmetric, SM gauge and fermionic
content have been   obtained.
 Such constructions were achieved by considering Type IIA
D6-branes wrapping at angles on a six dimensional torus in the presence of
orientifold planes and NS background fields.
The relevant observation there  is thus that quark doublets generations 
behave  differently  under  $U(2)$ transformations.

We should notice that many {\it Standard like models} found in the 
literature contain, apart from extra doublets, extra  
vector like triplets in the spectrum. The origin of these triplets is 
somewhat different and is related to the singularity structure. 
For instance, they usually appear 
due to  ``$mod$ $ N $''   identifications in  models with  $\IZ_N$ like 
singularities (see for instance D3-brane models in \cite{aiqu,bkl,bjl}).  
Once an extra  triplet appears, apart from those of the SM, then a 
corresponding $U(3)$ anti-triplet is again required for tadpole cancellation 
to occur.  
Interestingly enough, for  D3-brane models at $\IZ_N $ singularities, $\IZ_3$ 
like singularities are the only supersymmetric ones leading to 3 
(equivalent) generations
\cite{aiqu} and they always lead to  vector like  triplets.

In this note we attempt, by invoking similar arguments as in Ref.\cite{imr}
related to anomaly cancellation, to build models  as close as possible to 
the Standard Model,  in the framework of configurations  of
D3-branes stuck on a $\IR^6/\IZ_N$ and in the presence of D7- branes. 
By {\it ``as close as possible''} we mean the Standard  Model minimal content 
of 3 generations of quarks and leptons, without extra massless matter like, 
for  example, $SU(2)$ doublets or $SU(3)$ triplets.  

 In order for the spectrum to contain left quarks doublets as well as 
anti-doublets of $U(2)$,   the singularity  point is also required to  
be invariant under an orientation reversal (orientifold) action.
Moreover, in order to avoid  extra charged matter, due to orbifold-orientifold 
identifications,  non supersymmetric singularities with high values of $N$
must be considered.

Our analysis should be viewed  as a first step, in the spirit of a
bottom-up approach (see for instance \cite{aiqu}), in the
construction of a full string model. 

In fact, this partial  structure should
be further  embedded into  a globally  consistent string model.
 Depending on the features of the singularity, this could be achieved,
 for instance, by considering Type IIB
orientifold (a cristalographic singularity)  or  generically
F-theory compactifications \cite{aiqu,ftheory}. 
It is important to notice that many
phenomenological features will depend on the global structure.

An relevant  difference with respect to the approach in
\cite{aiqu}, that should become clear from discussion below  is
that, in our proposal, hypercharge necessarily involves $U(1)$
generators coming from 77 branes sectors if correct hypercharge 
(free of anomalies)  Standard model assignments are looked for. This means
that such groups must be gauged and not merely global symmetries.
Therefore, generically, states charged with respect to D7 groups
will carry hypercharge. For the full construction to be
consistent it should be ensured that hypercharge anomalies carried
by  D7 states could  be canceled or that, such states, could  be finally projected 
out.

We will not address this second step here, involving the full
construction of the D7 brane sector. In this sense, our
construction implies the identification of {\it necessary} conditions
for plausibly having just the Standard Model content in this context.
 Nevertheless, we will argue that there seems to be enough
 freedom in the 77 brane sector for achieving full consistency.

An extensive treatment of D3-branes at singularities was
presented in \cite{aiqu}. We will closely follow the notation used
there and  borrow some of the results. Let us recall some facts.

The states corresponding  to  a set of $n$ D3-branes stuck at a 
$\IR^6/\IZ_N$ singularity are obtained by keeping original 
fermionic and bosonic states invariant under the action of  
$\IZ_N$  generator $\theta$. Recall that  $\theta$ rotates string 
coordinates as well as Chan-Paton indices.
The latter can be achieved by a general twist matrix given by
\beqa
\label{gamma3}
\gamma_{\theta,3} = \diag (\id_{n_0}, e^{2\pi i/N} \id_{n_1},\ldots,
e^{2\pi i(N-1)/N} \id_{n_{N-1}})
\eeqa
where $\id_{n_i}$ is the $n_i\times n_i$ unit matrix, and $\sum_i
n_i=n$.
 Twist information can be encoded  into the vector
\beqa
\label{V3}
{\bf V}_{3} = \frac1N (0,\dots,0, 1,\dots,1, \dots,(N-1),\dots,(N-1) )
\eeqa

with ${n_0} \, 0's, {n_1}\,  1's $ etc.

  For instance, the   four fermionic states on the D3-brane world-volume,  are
described by Ramond  states $\lambda |s_1,s_2,s_3,s_4 \rangle$,
with $s_i=\pm \frac 12$ and $\sum_i s_i={\rm odd}$ where $\lambda $ is
a Chan-Paton factor. By convention we
choose  $s_4=-\frac 12$ to be left-handed fermions.
 $\IZ_N$ rotation on Fock  string states can be encoded in the vector
 $(a_1, a_2,a_3,a_4)$ with $a_1+a_2+a_3+a_4=0 \,\, {\rm mod} \,\, N$,
and it  is represented by
${\cal R}(\theta)  |s_1,s_2,s_3,s_4 \rangle= e^{2\pi i
a_{\alpha}s_{\alpha} /N} |s_1,s_2,s_3,s_4 \rangle$.  Invariant
fermionic states are given by

\beq
\label{33fs}
\lambda = e^{2\pi i a_{\alpha}/N} \gamma_{\theta,3} \lambda
\gamma_{\theta,3}^{-1}
\eeq
Similarly, the action on NS states, namely,  gauge bosons
$\lambda\psi^{\mu}_{-\frac 12}|0\rangle$, with $\mu$ along the
D3-brane,  or complex scalars given
by $\lambda\Psi_{-\frac12}^r|0\rangle$ (with $r=1,2,3$ labeling a complex
plane transverse to the D3-brane) can be encoded in a vector
$ (b_1,b_2,b_3,0)$ with $b_1=a_2+a_3$, $b_2=a_1+a_3$, $b_3=a_1+a_2$, 
where we have  included a fourth space-time world volume  coordinate
with $b_4=0$. Thus,  invariant NS states are given by
\beqa
\label{33NSs}
\lambda = e^{-2\pi i b_r/N}\ \gamma_{\theta,3}\ \lambda\
\gamma_{\theta,3}^{-1}
\eeqa
Spectra can be easily computed when Chan-Paton terms  are written in a 
Cartan Weyl basis (details can be found in\cite{afiv,aiq}). 
Namely, CP generators are organized into Cartan algebra generators
$\lambda_I = H_I$, $I=1,\cdots, {n}$ while charged generators are labeled by 
 $U(n)$  root vectors $\rho  _3= ({\underline  1,- 1,0\dots0})$ where  underlining indicates all possible  permutations.

Thus, (\ref{33fs}) and (\ref{33NSs}) select charged  generators satisfying 
 \begin{eqnarray}
\rho_3 \cdot V_3 &=& \frac{-a_{\alpha}}{N}  {\rm \, mod \,} {\IZ} \nonumber \\ 
\rho_3 \cdot V_3 &=& \frac{b_r}{N}  {\rm \, mod \, } {\IZ}
\label{33cw}
\end{eqnarray} 
for fermionic  and NS states respectively. Cartan generators are projected 
out whenever phases are non vanishing.
The resulting spectrum in the 33 sector reads
\beqa
{\rm Vectors} & \prod_{i=0}^{N-1} U(n_i)  \nonumber\\
{\rm Complex} \;\; {\rm Scalars} & \sum_{r=1}^3 \sum_{i=0}^{N-1}
(n_i,{\ov n}_{i-b_r}) \nonumber \\
{\rm Fermions} & \sum_{\alpha=1}^4 \sum_{i=0}^{N-1}  (n_i,{\ov
n}_{i+a_\alpha})
\label{33spec}
\eeqa
where sub-indices are  understood modulo $N$. Also, fundamental
(anti-fundamental) representations of $SU(n) $ carry unit ($-1 $)
charge with respect to the $U(1) $ factor in $ U(n) $ .

D7-branes are generically required in order to achieve cancellation of
 RR charges. Take, for instance,  $7_3 $ branes, orthogonal to complex
 coordinate $Y_3$ and containing a set of D3-branes as considered
 above and choose $ b_3={\rm even} $.
    The Chan-Paton embedding can be defined as 
\beqa
\gamma_{\theta,7_3} & = & \diag (\ \id_{u_0}, e^{2\pi i/N} \id_{u_1},\ldots,
e^{2\pi i(N-1)/N} \id_{u_{N-1}}) \quad \quad {\rm}\;\;
\eeqa 
with $\sum_i u_i=u$ and a corresponding {\it shift} vector ${\bf V}_{7_3}$ 
can be assigned as in eq. (\ref{V3}). 
The massless ${\bf 37_3}$  spectrum  
is then found from  the conditions 
\begin{eqnarray}
\rho^{37_3}\cdot V_{37_3} &=& -\oh \frac{b_3}{N}  {\rm \, mod \,} {\IZ} \nonumber \\ 
\rho^{37_3} \cdot V_{37_3} &=& \frac{(b_1+ b_2)}{2N}  {\rm \, mod \, } {\IZ}
\label{37cw}
\end{eqnarray}
for  left handed fermions and scalars respectively. Here 
 $V_{37_3}=(V_{3},V_{7_3})$ and $\rho  _{37_3}= 
({\underline  {1,0\dots0}};{\underline {- 1,0\dots0}})$
are $n+u$ dimensional vectors. 

The resulting 37+73  spectrum is 
\beqa
\begin{array}{llll}
 & & {\rm Fermions} &
\sum_{i=0}^{N-1}\, [\, (n_i,{\ov u}_{i+\frac 12 b_3}) +
(u_i,{\ov n}_{i+\frac 12 b_3}) \,] \\
 & & {\rm Complex}\;{\rm Scalars} &
\sum_{i=0}^{N-1}\, [\, (n_i,{\ov u}_{i-\frac 12 (b_1+b_2)}) +
(u_i,{\ov n}_{i-\frac 12 (b_1+b_2)}) \,] \\
\end{array}
\label{37spec}
\eeqa

Similar results are obtained for other D7$_r$-branes, transverse to the
$r^{th}$ complex plane, just by replacing $b_3 \to b_r$ etc.

Here we are concentrating in sectors containing the $n$  D3 at the 
singularity, since we pretend to place the SM on them. However, we must also 
 take  D7 branes into account. 
When non-compact configurations are considered  D7 branes are non dynamical 
and the corresponding 77 groups are global symmetries. 
Nevertheless, when the above configuration is embedded in a compact space, 77 
sector must be treated in an equal footing with the others. We will not 
address this computation here but we will comment on it below.

 The above  Chan-Paton twists, though  consistent with
 a $\IZ_N$ action, must be further constrained in order to ensure
 twisted RR fields charge cancellation.  
As is well known \cite{lr,iru,aiqu,abiu}  
for generic $ n_i, {u^ r}_i $ these are
equivalent to non-Abelian  $ SU(n_i) $ gauge anomaly
cancellation. Namely,
 \beqa
\sum_{\alpha=1}^4 (n_{i+a_\alpha}-n_{i-a_\alpha}) + \sum_{r=1}^3
(u^r_{i+\frac 12b_r}-u^r_{i-\frac 12 b_r}) = 0
\label{nonabanom}
\eeqa

Notice (see (\ref{33spec}-\ref{37spec})) that the term with a positive 
(negative) sign is the multiplicity of the ${\bf n}_i $ fundamental
(anti-fundamental) representation of  $ SU(n_i) $. Thus, the same number
 of fundamental and anti-fundamental $SU(n)$ representations must be present in
the spectrum. 
Equivalently, another way to read the above result is  that the net $ U(1)$
charge, for each $U(n)$,  must vanish since  a fundamental
representation carries charge 1  while anti-fundamental $-1$ .
Interestingly enough we find constraints even  when $n_i$ take specific 
values like  $n_i=0,1,2$ which would lead, respectively to, 
no gauge group at all or $U(1)$ or $U(2)$ where non-abelian anomalies are 
not expected.

Recall that, since  only fermions in  bi-fundamental representations  
of the form $(n_i,{\ov n}_{i+a_\alpha})$ appear in the 33 spectrum,  
we could have  $(3,{\ov 2})$ left quarks (here ${\ov 2}= 2_{-1}$) 
which are $U(2)$ anti-doublets (or  $({\ov 3},2)$) but not $(3, 2)$ 
doublets. Therefore,  if we managed to obtain 3 generations of 
left handed quarks, we would  {\it always} need six extra doublets as we 
have discussed above.
\footnote{ An alternative way, which deserves further 
investigation, to cancel extra doublets anomalies could be achieved 
by  turning on $B$ and $F$ fluxes in the lines recently suggested in 
Ref.\cite{urflux}}

 The  possibility of having different $U(2)$ transformations 
for left quark generations opens up when  $\IZ_N$ singularity
is placed  onto an orientifold plane. Indeed, orientifold identifications allow
 for the presence of $(n,n)$ and $({\bar n},{\bar n})$ as well.

 Invariance under an orientifold action $\Omega $ imposes further
constraints on twists considered above \cite{afiv}.
 In particular $\gamma_{\theta,3} =
({\tilde \gamma_{\theta,3}},{\tilde \gamma_{\theta,3}}^*)$.
 This twist leads to a replicated group with a spectrum invariant
under conjugation of representations. 
 Thus we see that ranks  $n_j=n_{-j}$ and factors $U(n_j)$ and
$U(n_{-j})$ are exchanged and must be identified in the quotient by $\Omega $.
 Similarly the fundamental representation  $n_j$  
 goes over to the anti-fundamental representation ${\ov n}_j$, and vice-versa.
 When the two entries of some bi-fundamental are charged
with respect to the same group in the quotient, the antisymmetric
combination must  be kept etc.. 
Again, an operative way to easily compute the spectrum  is to work in a 
Cartan-Weyl basis. In particular, equations (\ref{33cw}, \ref{37cw}) are still 
valid if a shift vector 

\beqa
\label{Vor}
{\bf V}_{3} = \frac1N (0,\dots,0, 1,\dots,1, \dots, P \dots,P ),
\eeqa
with $N=2P (2P+1)$,   is  assigned  to ${\tilde \gamma _3}$ 
(and similarly a ${\bf V}_{7_r}$ to ${\tilde \gamma _{7_3}}$  ) 
and by replacing $+$ and  $-$ signs by $\pm 1$ in root vectors. 

For instance  
$\rho  _{33}= ({\underline { \pm 1,\pm 1, 0\dots,0}})$ should be used. Notice 
that these  correspond to $SO(2n)$ charged generator roots, as expected, after 
orientifold projection.
We will not write down the generic spectrum but rather concentrate 
 on  specific examples in order to illustrate our discussion.
 
Let us look for an explicit realization of the above ideas.
Namely, we search for  SM gauge group  on D3 branes  at the $ZN$
orientifold singularity, in the presence of D7 branes, 
 with a basic structure of 3 left handed SM 
$SU(3) \times SU(2)\times  U(1)_Y $ quarks as
\begin{equation}
\label{leftq}
  3(3,2,1/6)= 2(3,\bar 2,1/6)_{1,-1}+ (3, 2,1/6)_{(1,1)}
\end{equation}

Subscripts indicate the charges corresponding to  $U(1) $ factors in 
$U(3) \times U(2)$ .
 
Let us first notice that it is natural to place both $SU(3)$ and $SU(2)$ 
gauge groups  on D3-branes. In fact, as can be seen already from the spectrum 
in eq.(\ref{37spec}), multiplicity of $7_r3 $ states is just one 
(due to second constraint in ( \ref{37cw})) and therefore,   
it is not possible 
to get 3 left handed quark  generations, for instance, by placing $SU(2)$ on a 
$77$ sector.
Since, as can be seen from eq.(\ref{33cw}),  
fermion multiplicity is given by the number equal twist eigenvalues,  we must  
have a  twist action on fermions  of the form
\begin{equation}
 (a_1, a_2,a_3,a_4)= (a,a,b,c) 
\label{ftwist}
\end{equation}
with $c=-(b+2a)$ $ mod$ $ N $. Moreover,  
$b\ne a \ne c$ $ mod$ $ N $ in order to avoid three identical generations and  
also $b \ne c$ $  mod$ $ N $.

A $(3,{\ov 2})$ is represented by a 33 root vector  
$= ({\underline {1, 0, 0}}, {\underline {-1, 0}},\dots)$,  while a $(3,{ 2})$ 
corresponds to  $= ({\underline {+,0,0}}, {\underline {+,0}}, \dots)$, 
where the 
first three entries correspond to $SU(3)$ and the two others to $SU(2)$.

Thus, we are lead to CP twists  of the form 
\begin{equation}
V= 1/N ( -\frac {a+b}2, -\frac {a+b}2,-\frac {a+b}2,\frac{a-b}2,
\frac{a-b}2, d_1,d_2, \dots )
\label{twistvec}
\end{equation}
It produces the desired  states in  eq.(\ref{leftq})  
if hypercharge is defined as  $Y= \frac{Q_3}6+ 0 \frac{Q_2}6+ \dots $

Entries $d_1,d_2$ etc. allow for the presence of extra $U(1)$ factors that
could be needed to accommodate the rest of the SM spectrum 
( one such factor is  added in model below).

We have stressed that $Q_2$ can not be part of hypercharge since
$U(2)$ doublets and antidoublets, carrying opposite such charge,  
must give the same $1/6$ hypercharge contribution. 
This would, necessarily, lead to include $U(1)$ charges originated in 
D7-brane groups in the definition of $Y$. The reason is that, 
as it can be checked, it is not possible to 
accommodate all right quarks and left leptons in 33 sector for any $N$. 
Some of them must necessarily come from $37_r$ sectors.
Since $Q_3$ normalization is already fixed in order to produce correct left 
quark assignements and $Q_2$ is not present in $Y$ 
correct charges for such states must include $U(1)$ generators from $77$  
sectors (otherwise right quarks from $37$ sectors would 
carry $Y=-1/6$ and/or leptons $Y=0$ hypercharge). 

Therefore D7-branes must  be embedded in a global compact
manifold for the groups to become  gauged. This differs 
from the approach in \cite{aiqu},  where hypercharge
appeared as the diagonal combination of D3-brane $U(1)$ groups.

Notice that  $b=-2a$  (mod $N$)  would lead to a supersymmetric singularity.
However, for the twist above, this value would produce extra $(3,1,-1/3)$ 
representations (coming from conjugate antisymmetric representations of 
$SU(3)$ ).
Since we are looking for  a fermionic content as close as possible to
 the Standard Model one,  this choice  for $b$ must be forbidden. 
 Let us point out that in supersymmetric models, extra doublets 
required by tadpole cancellation, could be interpreted as higgsinos and  
therefore, in such  cases, there is no real need to have different $U(2)$ 
behaviours for left quarks (if we allow for several Higgs fields). 
However, in this situation, $N=3$ is needed 
in order to have three generations and this choice alway leads to extra 
matter.     We conclude that it is {\it not possible} to obtain the, exactly, 
 Minimal Supersymmetric Standard Model from a D-brane at an orbifold or 
orientifold singularity. 

 Several other restrictions must be imposed on $a_{\alpha}$ and $N$ 
in order to avoid such kind of extra matter. 
For instance   $ a\ne -a,-b,-c $  in order to avoid $(\ov 3,2)$ 
states or $\frac{a-b}2 +d_i \ne - a_{\alpha}$ to forbid  anti-doublets etc. 
Similar constraints do appear when 73 sectors are included.  
 Observe that these restrictions are all  ``$modulo  N$  and, therefore,  
will require $N$   to be sufficiently large in order to forbidd 
identifications. 
 
In fact, it appears, as we indicate below,  that $N \ge 11$.

Let us see how all this works in   explicit  examples:

Consider the action on fermions given by odd twists $ a_{\alpha}$. Let us 
choose, for instance, $(a_1, a_2,a_3,a_4)= (1,1,-5,3)$.  Hence, the 
corresponding  action on scalars is  achieved by  
$(b_1, b_2,b_3 )= (-4,-4,2)$. 

The twist on Chan-Paton  factors in (\ref{twistvec}) becomes   
$ V= 1/N ( \dots,2,2,2,3,3,\dots )$ 
The constraints discussed above indicate  already that 
$N \ne 2,3, 4, 5, 6,8 $ or $10$. $N=7,9$ can also be discarded by a more 
carefull analysis of $37$ spectra.  We will choose $N=12$, for concreteness,  
and briefely discuss other possibilities afterwards. 

A $\IZ_{12}$ example:

 Even if generic features are shared with other $N$'s singularities, 
let us stress that 
$\IZ_{12}$ is a peculiar example since it corresponds to a 
cristalographic singularity. Let us briefely comment about this. 
  Classification of cristalographic singularities \cite{ek,dhvw} shows that 
 only some  $\IZ_{12}$,  non factorizable, singularities are possible and 
that factorizable one's are at most  of  $\IZ_{6}$ type. 
 Actually, notice that this corresponds to our case. Eventhough the action on 
fermions is given by a  $\IZ_{12}$ twist, the  lattice is
 defined by  the action on scalars  which  in the example  at hand  
is given by $\frac1{12}(-4,-4,2) =\frac1{6}(-2,-2,1)$,  namely,  a product of 
three hexagonal ( $SU(3)$)   lattices.   
 Therefore, it would be possible, for instance, 
to achieve a full consistent orientifold compactification 
In particular, the possibility of having large extra dimensions 
\cite{anto,larged}  in order to lower the string scale, which is of 
phenomenological interest here since models are non supersymmetric,
  is open in this context. 

Recall that  $\sum b_i = even$  as is required by modular 
invariance in the closed, torus, sector. Nevertheless, 
since the singularity is  not supersymmetric, tachyons in the closed string 
sector must be taken care of  (see for instance  \cite{fh})). 

As explained above, in order to ensure twisted tadpole cancellation, 
we must first write down the  generic spectrum and  then find the 
conditions for  it to be free of anomalies. 
After that we may choose specific values  for  the number of  given CP twist 
eigenvalues  ($n_i's, u_i's $ etc.) satisfying such requirements, 
in order to build a specific model.

Thus, for the 33 sector we define a generic twist  
\begin{eqnarray}\label{vgeneral}  
 V_{3} & =& \frac1N (1,...1,2,...2,...6,...6)  
 \end{eqnarray}

Where there are $n_i$ entries equal to i on the
 D3-branes ($n_0$ is chosen to vanish) 
and similar vectors for $D7_1$,$D7_2$ and $D7_3$-branes with 
with $u_i,v_i$ and  $w_i $ entries.

The spectrum reads

\begin{eqnarray}  
\label{33Z12}  
 & & Sector 33 \\ \nonumber
& & 2[(n_1,\ov 3)+(3, \ov 2) + (2, \ov n_4) + (n_4, \ov n_5) +
 (n_5, \ov n_6) + (n_5,n_6)] + \\\nonumber
 & & (n_1, \ov n_4) + (3, \ov n_5) + (2,\ov n_6) + (2, n_6) +
 (n_4, n_5) + (\ov n_1,\ov 3) + \\ \nonumber& & 
 (\ov n_1, \ov n_6) + (\ov  n_1, n_6) + (n_1, n_4) + (3, 2) + 
(\ov 3, \ov n_5)
+ (\ov 2, \ov n_4)\nonumber
\end{eqnarray} 
\begin{eqnarray}
\label{373Z12} 
 & & Sector 37_3 \\\nonumber
& & (n_1, \ov w_2)+(3,\ov w_3)+(2,\ov w_4)+(n_4,\ov w_5)+
(n_5,\ov w_6)+(n_5,w_6)+\\ \nonumber
& & (\ov 3,w_1)+(\ov 2,w_2)+(\ov n_4,w_3)+(\ov n_5,w_4)+(n_6,w_5)+(\ov n_6,w_5)
\end{eqnarray} 
\begin{eqnarray}\label{371Z12} 
& & Sector 37_1 \\\nonumber
& & (n_1,u_1)+(\ov n_5,\ov u_5)+(\ov n_4,\ov u_6)+(\ov n_4,u_6)
+(n_6,\ov u_4)+(\ov n_6,\ov u_4)+ \\ \nonumber 
& &
(\ov n_1,u_3)+(\ov 3,u_4)+(\ov 2,u_5)+(2,\ov u_1)+(n_4,\ov u_2)+(n_5,\ov u_3)
\end{eqnarray} 
and similarly  for sector $37_2$ by replacing $u \rightarrow v$. 
For the sake of clarity we have already chosen $n_2=3$ and $n_3=2$.

The twisted tadpole cancellation requirements read 
\begin{eqnarray}
\label{anomcanc}
-3n_1+3n_3+w_3-w_1-u_4-v_4=0\\
-n2+n_4+2n_6+w_4-w_2+u_1+v_1-u_5-v_5=0\\
n_2+2n_4-2n_6+w_2+u_1+v_1-u_3-v_3=0\\
3n_5-3n_3+w_5-w_3-2u_6-2v_6+u_2+v_2=0\\
-n_4+4n_6-2n_2+2w_6-w_4+u_3+v_3-u_5-v_5=0
\end{eqnarray}
\bigskip

Clearly    $n_5=n_4=n_6=w_2=w_3=u_5=v_5=0$ in order 
to avoid extra triplets or anti-doublets and $n_1=1$. 
Interestingly enough, since 
$n_2=3$ and $n_3=2$, we find that  first equation becomes 
$ w_1+u_4+v_4=3$ 
telling us, as expected, that 3 anti-triplets must be provided by 
$37_r$ sectors. Similarly second equation indicates that 
$ w_4 +u_1+v_1=3$ doublets must come from such sectors.
  $3+u_1+v_1=u_3+v_3$ requires the total number of 
$1$ and $\ov 1$ to be the same. 
The other two equations come from cancellation of, generic,  
$SU(n_4)\times SU(n_5)$ anomalies even though here, $n_4=n_5=0$.

By identifying the two anti-triplets $2(n_1,\ov 3)$ 
 in 33 sector with $U_R$ quarks (and therefore $(\ov n_1,\ov 3)$ with $D_R$) 
and placing the third one in $37_3$ sector ( $w_1=1$)  we must define 
the hypercharge as
\begin{eqnarray}  
\label{hyperc}  
Y &=&  \frac{Q_3}6 -\frac{Q_1}2+\frac{Q_a}2+ \frac{Q_b}2+\frac{ Q_c}2 +
 {Q_{neutral}^{7}}  
\nonumber  
\end{eqnarray}  

with  
\begin{eqnarray}  
\label{abcharge}  
Q_a &=&   {Q_{4}^{7_1}} + Q_{4}^{7_2}  - Q_{1}^{7_3}\\
Q_b & =&  Q_{4}^{7_3}\\
Q_c &= &  Q_{3}^{7_2}  + Q_{3}^{7_1}  
 \nonumber  
\end{eqnarray}  
where the sub-indices indicate the corresponding Chan-Paton twist 
eigenvalue.
We have summarized in   $Q_{neutral}^{7}$ the possibility of including  other 
charges from 77 sectors. Even if SM massless fermions carry no such charge 
massless scalars could be charged.

Since the number of right handed leptons is  given by  $u_3+v_3 $,
then  $u_1=v_1=0$ must be imposed.

We see that, due to the symmetry between  $7_1$ and $7_2$ 
branes (see \ref{37cw}) there is still certain freedom 
to place some states in one or another sector leading to same spectrum.

As an example, let us choose  to place three ($u_3=3$)  right leptons in 
$37_1$ sector and two $D_R$ quarks ($v_4=2$) $37_2$.
Therefore, vector shifts 
  
\begin{eqnarray}\label{vz12}  
 V_{3} & =& \frac1N (1,2,2,2,3,3)\\  
  V_{7_1} & =& \frac1N (2,\dots 3,3,3,6,\dots 6)\\ 
  V_{7_2} & =& \frac1N ((2,\dots 2,4,4,6,\dots 6)\\
V_{7_3} & =& \frac1N (1,4,4,4,5\dots 5,6,6,6)  
\end{eqnarray}
with 
$w_5+u_2+v_2=6+ 2(u_6+v_6)$,
  lead  to  $SU(3)\times SU(2) \times U(1)_Y$ spectrum

\begin{eqnarray}
\label{SMs}
& &  2(3,\ov 2,\frac 16) +(3, 2,\frac 16) + 3(\ov 3, 1,\frac {-2}3 )+
3(\ov 3, 1,\frac {1}3) \\ 
& & 3(1,2,\frac {-1}2)+ 3(1,1,1)\nonumber
\end{eqnarray}

 Results are summarized in Table 1
\begin{table}[htb] \footnotesize
\renewcommand{\arraystretch}{1.25}
\begin{center}
\begin{tabular}{|c|c|c|c|c|c|c|c|c|c|}
\hline  Matter fields  &
 Sector &   &  $Q_3$  & $Q_2 $ & $Q_1$ & $Q_a $ & $Q_b$&  $Q_c $ & Y \\
\hline\hline $Q_L$ & $(33)$ &  $2(3,{\ov 2})$ & 1  & -1 & 0 & 0&0 & 0&1/6 \\
\hline  $q_L$ & $(33)$  &  $( 3,2)$ &  1  & 1  & 0  & 0 &0&0 & 1/6 \\
\hline   $U_R$ &  (33) &  $2( {\bar 3},1)$ &  -1  & 0  & 1  & 0&0& 0  & -2/3 \\
\hline         &  $(37_3)$ &  $ ( {\bar 3},1)$ &  -1  & 0  & 0 & -1 &
0&0 & -2/3\\
\hline  $D_R$ & 33  &  $ ({\bar 3},1)$ &  -1  & 0  & -1  & 0 &0& 0& 1/3 \\
\hline     &$ 37_2 $ &  $2( {\bar 3},1)$ &  -1  & 0  & 0  & 1&0 & 0& 1/3 \\
\hline  $ L$ & $ 37_3$    &  $ 3(1,2)$ &  0   & 1   & 0  & 0 & -1 & 0 & -1/2 \\
\hline  $E_R$  & $37_1$   &  $3(1,1)$ &  0  & 0  & -1  & 0&0&1 & 1   \\
\hline 
\end{tabular}
\end{center} \caption{ Standard model spectrum and $U(1)$ charges
\label{tabpsm} }
\end{table}

Notice that when above $U(1)$ charges are global symmetries they 
can  be given a familiar physical meaning (see \cite{imr} and 
\cite{afiu} for  similar observations).  
In particular,  
$ Q_3= 3B$ where $B$ is the baryon number and $L=-Q_c-Q_b$ is 
the lepton number. 
 Also $ I_R= \frac{1}{2}(Q_1-Q_a) $  corresponds to $SU(2)_R$  weak isospin 
in left-right models etc. 
Recall, however, that at  this level all such symmetries are actually local 
symmetries.

Several comments are in order.
A relevant observation  is that, 
correct hypercharge assignments, ensuring that above 
$SU(3)\times SU(2)\times U(1)_Y$ spectrum is anomaly free,
unambiguously (up to some signs depending on which sector we chose to 
place right up or down quarks) define hypercharge  $Y$ above (\ref{hyperc}).
Therefore,  $Y$  must involve generators of 77 sectors Abelian groups.

As stressed before this appears to be an unavoidable fact also for other 
$N$'s. There is no way to obtain the SM content from only the 33 sector 
without including extra matter.  

A consequence  is that $7_r7_r$ states, 
which are singlets under $SU(3)\times SU(2)$,  will generically carry 
hypercharge. Hence, mixed anomalies 
of $Y$ with  $7_r7_r$ groups should vanish in order for $Y$ to be 
truly free of anomalies. 

A complete study of this fact is out of the scope of this note where 
we are looking  for a set of necessary conditions to achieve 
a {\it minimal} SM content.
 Notice, however, a somewhat related fact.  Quite plausibly, Wilson lines (WL) 
could be introduced in order to break $U(N)$ factors into products of abelian 
groups in 77 sectors (some of which will become massive). Thus, the 
dimension of 77 representations in 37 sectors will  become true  
multiplicities for SM charged states as written down in eq.(\ref{SMs}).   
Moreover, in this procedure, by suitably choosing WL,  
77 states carrying hypercharge could be completely projected out of 
the  spectrum thus leaving, at most,  a certain number of 
SM singlets $(1,1,0)$ and/or  hidden matter. 

In order to indicate how this could work,  let us turn on a discrete 
Wilson line on second complex plane \cite{aiqu,afiv,cuw}.
This WL must be embedded as a twist on $7_1$ and $7_3$ branes. For instance, 
we can choose the twist on  $7_1$ -branes to be  
\begin{eqnarray}
 W_{7_1} & =& \frac1{12} (3,\dots, 4\dots, 5\dots,6,\dots; 0,1,2; 3\dots)
\end{eqnarray}
where  the first (last)  entries correspond to $l_i$ $i's$  $i\ne 0,1,2$ 
 ( $m_i$ $i's$  $i\ne 0,1,2$ ) with 
 $\sum l_i=u_2$ ( $\sum m_i=u_6$). 
Such a WL breaks $U(3)_3$ to $U(1)^3$   and also projects out all 
 $Y$ charged states 
 $(u_2,\ov u_3)_{-1/2}+ (u_2, u_3)_{1/2}$ as desired.
Notice that there is plenty  of freedom for choosing  WL producing such 
effect (moreover, in this particular case we could have chosen $u_2=0$ 
without even needing to introduce a Wilson line).

We can proceed similarly with the  twist on $7_3$  branes. 
The same goal can be achieved on $7_27_2$ sectors by adding a WL on 
first (or third) plane with a corresponding action on ,$7_2$ and $7_3$ 
($7_2$ and $7_1$)-branes.

Unfortunately this is not the whole story. $7_r$ branes contain  other four  
fixed points, apart from the origin. Twisted tadpoles are expected at 
such points and therefore extra D3'-branes must be placed there in order to 
cancel them. Again, among  $3'7_r$ states, some of them will generically carry
 hypercharge.
 Since these are not fixed by the orientifold action, spectra and 
tadpole cancellation conditions will be of the form discussed in 
eq. (\ref{33spec}-  \ref{nonabanom}). 
Moreover, such points will generically feel the presence of the Wilson lines.

Due to the freedom in choosing both Wilson lines and twists  on 
D3'-branes we expect to be able to achieve such cancellation at each point, 
 by forbidding (hyper)-charged $3'7_r$ states.
For instance, fixed point $P=(0,+1,0)$ will feel a twist on $7_1$ branes of 
the  form 
 \begin{eqnarray}
V_P= ( V_{7_1}+W_{7_1},- V_{7_1}-W_{7_1})
\end{eqnarray}
Tadpole cancellation at such point eq. (\ref{nonabanom}) leads to twelve ($N=12$) 
 equations.
It can be checked that, in order to avoid (hyper) charged  states 
$n'_1=n'_2=n'_3= n'_5=n'_6=n'_7=0$. However, even if this is very 
restrictive there 
is enough freedom in choosing $n'_4, n'_8,n'_9,n'_{10}, n'_{11}$, $u_2$ and 
$u_6$ WL entries above to solve the equations system.

Thus, by suitable introduction of Wilson lines, all (hyper)-charged 77 states 
could be   projected out and relevant non-abelian groups broken 
down to Abel-Ian factors by leaving just gauge bosons and 
chiral fermions of the Standard Model (plus, presumably,  SM singlets).
Full consistent compactification will still require cancellation of untwisted 
tadpoles and twisted tadpoles proportional to inverse volume terms, not 
present in the infinite volume limit. In principle this could be achieved 
by  following similar steps as in the compact models constructed in 
\cite{aiqu} and \cite{ru,kr} for  example.  Certainly anti-branes will be 
needed etc.

Here we have concentrated in the fermionic spectrum. However, above twists 
 predict also the presence of scalars, as can be seen from (\ref{33spec}) and  
\ref{37spec}, charged under SM group. For instance, 
scalars $2(3,1,-1/3)$ will always be present with a generic twist as in 
\ref{twistvec} in 33 sector etc.
  
In particular, the spectrum for the model presented above contains 
 $ 2(3,1,\frac {-1}3) +3(1, 2,\frac 12))$  in 33 sector and   
$(u_2+v_2) [(1, 2,*)+(1, 1,*)] +3( 3, 1,\frac {-1}3 )+ 2(1,2,-1/2)$ from 
$37_1 + 37_2$ sectors and 
$(1, 2,1/2) +w_5 [(1,1,*)+ (1, 2,*)]+6 ( 3, 1,*) $ from $37_3$  sector. 
We have indicated  by $*$ that $Y$ charge depends on the definition of 
$ Q_{neutral}^{7}$ in (\ref{hyperc}). 
 
We will not address a phenomenological study here. Let us 
say that we expect such scalars to become generically massive, since no 
symmetry that would keep them massless is operating here.  They will 
contribute to the fermionic mass structure. 
Notice also  that, among scalars,  there are doublets with 
correct hypercharges to be identified with Higgs fields, which are 
required to break electroweak symmetry.

Twists defined above (\ref{vz12}) correspond to a so called 
{\it with vector structure twist} \cite{intri}
 (roughly speaking $\gamma ^N=1$). 
However, twists   {\it without vector structure}( $\gamma ^N=-1$) 
could  also be considered. 
We  have looked at $\IZ_{12}$ examples with twist 
$a_\alpha\equiv  1/12(-1-1-2 4)$  (thus $b=1/{12}(-3,-3,-2)$) 
and generic CP embeddings  
$V_3= 1/{24}(1\dots 1,3\dots 3,\dots , 11\dots 11)$. In such cases,  
twisted tadpole  cancellation conditions appear to be much  
stronger than above (due essentially to the presence of antisymmetric 
representations,  now allowed by $\a_{\alpha}$'s) and, as a consequence, 
 extra matter is required  in order to satisfy them.


Let us conclude with  a brief summary of our results and some observations.
We have identified a set of generic necessary conditions which appear to be 
required if a {\it minimal} Standard Model content is looked for in the 
context of  D-branes at  $\IR^6/\IZ_N$ singularities. We have concluded 
that,   $\IZ_N$ must be an orientifold fixed point, $N$ must be large enough 
 ($N \ge 11$ ) and singularity must be non-supersymmetric. Moreover, we have 
 shown that hypercharge  involves combinations  of $U(1)$ generators 
which must include, necessarily,  Abelian factors from 77 sectors.  
We argued that extra 77 or 73' states carrying hypercharge 
could be projected out, for instance, by suitable introduction of Wilson lines.
An explicit $Z_{12}$ example was presented to illustrate such issues.
Further examples, with higher order singularities,  can be treated in the same 
 way
\footnote{The spectrum can be obtained from \ref{33Z12}-\ref{371Z12}  
by a straightforward generalization taking care of $mod$ $N$ identifications}. 
However, since such cases are not cristalographic, they should be 
embedded in a more complicated generic Calabi- Yau compact space.

Let us emphasize that this is just a first step in the construction of a 
fully consistent model.  
In particular, since singularities are non supersymmetric, closed string 
tachyons will be generically present. 
They could  completely ruin the viability of these non-susy singularities or, 
more hopefully, they could be the signal of a transition towards more stable 
configurations as in the situations analised in \cite{ctachy}. Let us notice 
that, depending on the $\theta $ action on scalars,$ (b_1,b_2,b_3)$, 
 just closed  twisted tachyons, associated to orbifold fixed points can be 
present or  also closed tachyons propagating in the bulk 
(this is the case for the $\IZ_{12}$ example above).

Another relevant point refers to the possible couplings of closed 
 twisted RR fields to $U(1)$ field strengths \cite{iru,sagnan}. Such 
couplings, through a generalized Green -Schwarz \cite{gs} mechanism, 
 will ensure cancellation of $U(1)$ charges combinations possessing triangle 
anomalies. Corresponding $U(1)$ become massive. 
This should be the case, for instance, for
$Q_1$ or $Q_2$  anomalous combinations. 

Notice, however, that RR fields could couple to some (or all) 
anomaly free combinations, and render the corresponding gauge bosons massive. 
Further investigation of couplings  is required in order to ensure that 
$Y$ remains effectively a gauge symmetry.
Assuming that this is indeed the case, let us recall  that, for other  
$U(1)$'s  becoming massive, original gauge 
symmetries remain as global symmetries \cite{u1s} 
in the effective field theory. 
This leads to relevant phenomenological consequences. For instance, 
conservation of baryon number will protect proton from decaying  etc. 
As stressed, we will not pursue a  phenomenological study of models 
presented above.
Nevertheless, notice that the kind of analysis made in Ref.\cite{imr} 
(see also \cite{cim}), in the context of D6-branes at angles referring to 
such global symmetries,  can be paralleled here.  
In fact, this  appears to be generic to SM built up from D-branes where 
originally $U(n)$ factors rather than $SU(n)$ are present.

 \centerline{\bf Acknowledgements}
We are grateful to D. Badagnani and especially to 
 L.E. Ib\'a\~nez  and A. Uranga for stimulating  discussions and suggestions.
G.A. thanks U.A.M for hospitality. 
G.A work is partially supported by ANPCyT grant 03-03403.


\newpage

\begin{thebibliography}{99}
%
%
\bibitem{imr}
 L.E. Ibanez, F. Marchesano, R. Rabadan,JHEP 0111 (2001) 002 ( hep-th/0105155)
%
\bibitem{pheno}
 Some possible reviews on string phenomenology 
with reference to the original literature are: 
\\
F. Quevedo, hep-ph/9707434; hep-th/9603074 ;\\
K. Dienes, hep-ph/0004129; hep-th/9602045 ;\\
J.D. Lykken, hep-ph/9903026; hep-th/9607144 ;\\
M. Dine, hep-th/0003175;\\
G. Aldazabal, hep-th/9507162 ;\\
L.E. Ib\'a\~nez, hep-ph/9911499;hep-ph/9804238;hep-th/9505098;\\
Z. Kakushadze and S.-H.H. Tye, hep-th/9512155;\\
I. Antoniadis, hep-th/0102202;\\
E. Dudas, hep-ph/0006190;\\
D.Bailin, G. Kraniotis, A. Love, 
Searching for String theories of the Standard Model,[ hep-th/0108127]. 
%
\bibitem{guts}
See for instance:\\
 A. Font, L. Ibanez, F. Quevedo, Higher level Kac-Moody string models  
and their phenomenological implications, Nucl.Phys.B345:389-430,1990;\\
 G. Aldazabal, A. Font, L. E. Ibanez, A.M. Uranga, String Guts, 
 Nucl.Phys.B452:3-44,1995  [hep-th 9410206];\\
J. Erler, Asymmetric orbifolds and higher level models,  
Nucl.Phys.B475:597-626,1996 [hep-th/9602032]; \\
K. Dienes, J. March-Russell, Realizing  higher level gauge symmetries in 
string theory: New embeddings for String GUTs,
 Nucl.Phys.B479:113-172,1996 [hep-th/9604112].
%
\bibitem{aiqu}
G.~Aldazabal, L.~E.~Ib\'a\~nez, F.~Quevedo, A.~M.~Uranga,
D-branes at singularities: A Bottom up approach to the string embedding
 of the standard model,
JHEP 0008:002,2000 [hep-th/0005067].  
%
\bibitem{bkl}
 D. Bailin, G.V. Kraniotis, A. Love, Supersymmetric Standard Models on 
D-branes, Phys.Lett.B502:209-215,2001
\bibitem{bjl}
D. Berenstein, V. Jejjala and R.G. Leigh,
The Standard Model on a D-brane,
hep-ph/0105042.

\bibitem{ftheory}
C.~Vafa, \NPB{469}{96}{403}, hep-th/9602022.
%
\bibitem{afiv}
G.~Aldazabal, A.~Font, L.~E.~Ib{\'a}{\~n}ez, G.~Violero, \NPB{536}{98}{29},
[hep-th/9804026].
%
\bibitem{aiq}
G. Aldazabal,L. Ibanez, F. Quevedo
Standard-like models with broken supersymmetry from Type I string vacua, 
JHEP 0001:031,2000 [hep-th/9909172].
%
\bibitem{abiu}
G.~Aldazabal, D.~Badagnani, L.~E.~Ib{\'a}{\~n}ez, A.~M.~Uranga, JHEP
9906(1999)031 [hep-th/9904071].
%
\bibitem{lr}
Brane, anomalies,bendings and tadpoles,
By Robert G. Leigh, Moshe Rozali, Phys.Rev.D59:026004,1999 [hep-th/9807082]
%
\bibitem{iru}
L.~E.~Ib\'a\~nez, R.~Rabad\'an, A.~M.~Uranga,
Anomalous U(1)'s in type I and type IIB D = 4, N=1 string vacua,
Nucl.Phys. B542 (1999) 112-138, hep-th/9808139.
%
\bibitem{urflux}
A. M. Uranga, D-brane fluxes and chirality, 
A M. Uranga [hep-th/0201221] 
%
\bibitem{ek}
Jens Erler, Albrecht Klemm,
 Comment on the generation number in orbifold compactifications, 
 Commun.Math.Phys.153:579-604,1993,  [hep-th/9207111] 
%
\bibitem{dhvw} L. Dixon, J. Harvey, C. Vafa and E. Witten, {\em  Strings on 
orbifolds}, {\sl Nucl. Phys.} {\bf B261} (1985) 678 and 
{\sl Nucl. Phys.} {\bf B274} (1986) 285
%
\bibitem{anto}
I.~Antoniadis, \PLB{246}{90}{377}
%
\bibitem{larged}
See {\em e.g.} N.~Arkani-Hamed, S.~Dimopoulos, G.~Dvali, \PLB{429}{98}{263},
[hep-ph/9803315]; K.~R.~Dienes, E.~Dudas, T.~Gherghetta, \PLB{436}{98}{55},
[hep-ph/9803466]; I.~Antoniadis, N.~Arkani-Hamed, S.~Dimopoulos, G.~Dvali,
\PLB{436}{98}{257} [ hep-ph/9804398].
%
\bibitem{fh}
A. Font, A. Hernandez, Non-supersymmteric orbifolds [hep-th/0202057].
%
\bibitem{afiu}
 G. Aldazabal, S. Franco, Luis E. Ibanez, R. Rabadan, A.M. Uranga,
Intersecting brane worlds, JHEP 0102:047,2001 [hep-ph/0011132 ].
%
\bibitem{cuw}
M. Cvetic, Angel M. Uranga, Jing Wang, Discrete Wilson lines 
in  N=1 D = 4  Type IIB orientifolds:A systematica exploration for  Z(6) 
orientifold, 
Nucl.Phys.B595:63-92,2001 [hep-th/0010091]. 
%
\bibitem{ru}
R. Rabadan, A. M. Uranga, Type IIB Orientifolds without Untwisted Tadpoles, 
and non-BPS D-branes, JHEP 0101 (2001) 029 [hep-th/0009135] 
%
\bibitem{kr} 
M. Klein, R. Rabadan, D=4, N=1 orientifolds with vector structure,
Nucl.Phys. B596 (2001) 197-230,[hep-th/0007087]
%
\bibitem{intri}
K.~Intriligator, `RG fixed points in six-dimensions via branes at orbifold
singularities', \NPB{496}{97}{177} [hep-th/9702038]; J.~D.~Blum,
K.~Intriligator, ``Consistency conditions for branes at
orbifold singularities'', \NPB{506}{97}{223} [hep-th/9705030].
%
\bibitem{ctachy} 
 A. Adams, J. Polchinski , E. Silverstein,
 Dont'panic! Closed string tachyions in ALE space-times,
 JHEP 0110:029,2001 [hep-th/0108075]. 
%
\bibitem{gs}
M. Green and J.H. Schwarz, \PLB{149}{84}{117}.
%
\bibitem{sagnan}
A.~Sagnotti,
A Note on the Green-Schwarz mechanism in open string theories,
Phys. Lett. B294 (1992) 196, hep-th/9210127.
%
\bibitem{u1s}
L. Ibanez, F. Quevedo. 
Anomalous  U(1)'S and proton stability in brane models,  
JHEP 9910:001,1999 [ hep-ph/9908305];\\ 
C.P. Burgess, L. Ibanez, F. Quevedo, Strings at intermediate scale, or is 
the Fermi sacle dual to the Planck scale?,  
 Phys.Lett.B447:257-265,1999 [hep-ph/9810535] 
\bibitem{cim}
D. Cremades, L.E. Ibanez, F. Marchesano, 
 SUSY Quivers, Intersecting Branes and the Modest Hierarchy Problem
[ hep-th/0201205]    
%
\end{thebibliography}
\end{document}